\documentclass[runningheads]{llncs}
\usepackage[utf8]{inputenc}
\usepackage{textcomp}
\usepackage[table]{xcolor}

\usepackage[T1]{fontenc}

\usepackage[numbers,sort&compress]{natbib}
\usepackage{amsmath,amssymb,amsfonts}

\usepackage{url}
\usepackage{booktabs}
\usepackage{array}
\usepackage{tabularx}

\usepackage{amsmath}
\usepackage{algorithm}
\usepackage{algpseudocode}
\usepackage{graphicx}
\usepackage{textcomp}
\usepackage{xcolor}
\usepackage{tabularx}
\usepackage{subcaption}

\usepackage{algorithm}
\usepackage{algpseudocode} 
\usepackage{amsmath}       
\usepackage{graphicx}      
\usepackage{caption}       
\usepackage{multirow}
\usepackage[most]{tcolorbox}
\usepackage{underscore}
\usepackage{times}
\usepackage{latexsym}
\usepackage{booktabs}
\usepackage{atbegshi}
\usepackage[T1]{fontenc}
\usepackage[utf8]{inputenc}
\usepackage{microtype}
\usepackage{inconsolata}
\usepackage{pifont}
\usepackage{adjustbox}

\begin{document}
\title{Automated Visualization Code Synthesis via Multi-Path Reasoning and Feedback-Driven Optimizations}
\titlerunning{VisPath}

\author{Wonduk Seo\inst{1} \and
Daye Kang\inst{1,2*} \and
Hyunjin An\inst{1}\thanks{denotes co-second authors} \and
Taehan Kim\inst{3} \and
Soohyuk Cho\inst{4} \and
Seungyong Lee \and
Minhyeong Yu\inst{1} \and
Jian Park\inst{5} \and
Yi Bu\inst{6}\thanks{denotes corresponding authors} \and
Seunghyun Lee\inst{1}\textsuperscript{\scriptsize **}
}
\authorrunning{Seo et al.}
\institute{AI Research, Enhans, Seoul, South Korea \\ \and
Innovation \& Technology, KAIST, Daejeon, South Korea\\ \and
Department of Computer Science, University of California, Berkeley, CA, United States \and
Department of Electrical and Computer Engineering, Princeton University, Princeton, NJ, United States \and
Department of Data Science, Fudan University, Shanghai, China \and
Department of Information Management, Peking University, Beijing, China\\
\email{\{wonduk, hyunjin, minhyeong, seunghyun\}@enhans.ai}, \email{daye.kang@kaist.ac.kr}, \email{terry.kim@berkeley.edu}, \email{
soohyuk.cho@princeton.edu}, \email{buyi@pku.edu.cn}
}
\maketitle              % typeset the header of the contribution
\begin{abstract}
Large Language Models (LLMs) have become a cornerstone for automated visualization code generation, enabling users to create charts through natural language instructions. Despite improvements from techniques like few-shot prompting and query expansion, existing methods often struggle when requests are underspecified in actionable details (e.g., data preprocessing assumptions, solver or library choices, etc.), frequently necessitating manual intervention. To overcome these limitations, we propose \textit{VisPath: a Multi-Path Reasoning and Feedback-Driven Optimization Framework for Visualization Code Generation}. \textit{VisPath} handles underspecified queries through structured, multi-stage processing. It begins by using \emph{Chain-of-Thought} (CoT) prompting to reformulate the initial user input, generating multiple extended queries in parallel to surface alternative plausible concretizations of the request. These queries then generate candidate visualization scripts, which are executed to produce diverse images. By assessing the visual quality and correctness of each output, \textit{VisPath} generates targeted feedback that is aggregated to synthesize an optimal final result. Extensive experiments on \emph{MatPlotBench} and \emph{Qwen-Agent Code Interpreter Benchmark} show that VisPath outperforms state-of-the-art methods, providing a more reliable framework for AI-driven visualization generation.

\end{abstract}
\section{Introduction}

Data visualization has long been an essential tool in data analysis and scientific research, enabling users to uncover patterns and relationships in complex datasets~\cite{vondrick2013hoggles,demiralp2017foresight,unwin2020data,li2024visualization}. Traditionally, creating visualizations requires manually writing code using libraries such as Matplotlib, Seaborn, or D3.js~\cite{barrett2005matplotlib,bisong2019matplotlib,zhu2013data}. These approaches demand programming expertise and significant effort to craft effective visual representations, which can be a barrier for many users~\cite{bresciani2015pitfalls,saket2018task,sharif2024understanding}. As datasets continue to grow in size and complexity, researchers have explored ways to automate visualization generation, aiming to make the process more efficient and accessible~\cite{wang2015big,dibia2019data2vis,qian2021learning}.

% In response to this challenge,
Large Language Models (LLMs) have emerged as a promising solution for simplifying visualization creation~\cite{wang2023data,han2023chartllama,xie2024haichart,wen2025exploring}. By translating natural language instructions into executable code, LLM-based systems eliminate the need for extensive programming knowledge, allowing users to generate visualizations more intuitively~\cite{xiao2023let,ge2023automatic,zhang2024gpt}.
More recently, Chat2VIS~\cite{maddigan2023Chat2VIS} and MatPlotAgent~\cite{yang2024matplotagent} have been introduced to improve automated visualization code generation. Specifically, Chat2VIS follows a prefix-based approach, guiding LLMs to generate visualization code consistently; and MatPlotAgent expands the query before code generation. However, these methods face several limitations: (1) they generate code in a single-path manner, limiting exploration of alternative solutions, and are unable to recover when generating erroneous code; (2) they rely on predefined structures or examples, which restrict adaptability to ambiguous or unconventional user queries; and (3) a fundamental limitation is their inability to aggregate and synthesize multi-dimensional feedback. Without a mechanism to retrieve outputs that reflect diverse possibilities, they 
face difficulties in capturing the intricate details required for visualizations that are both functionally precise and contextually relevant.

To address underspecified visualization requests, we propose VisPath, a Multi-Path Reasoning and Feedback-Driven Optimization framework. VisPath samples multiple plausible concretizations of implicit implementation choices, executes candidate scripts, and leverages visual feedback to synthesize a robust final program.

Rather than directly translating user input into code, VisPath systematically accounts for both explicit requirements and implicit necessities to produce visualizations that are correct and insightful. It generates multiple reasoning paths that interpret user intent from different perspectives, producing structured blueprints that are converted into visualization scripts via Chain-of-Thought prompting. These candidates are then evaluated by a Vision-Language Model (VLM) for accuracy, clarity, and alignment with the intended message, and refined by a synthesis agent to optimize reliability and impact.

Experiments show that VisPath improves plot-level correctness and executability over prompting and agent-based baselines. Ablations attribute the gains to multi-path exploration and feedback integration, demonstrating stronger intent capture, higher execution reliability, and fewer errors—making visualization code generation more accessible for business intelligence, scientific research, and automated reporting.

\section{Related Work}
Numerous methods have been applied for Text-to-Visualization (Text2Vis) generation, which has significantly evolved over the years, adapting to new paradigms in data visualization and natural language processing~\cite{dibia2019data2vis,wu2022nuwa,chen2022type,chen2022nl2interface,rashid2022text2chart,zhang2024chartifytext}. Early approaches such as Voyager~\cite{wongsuphasawat2015voyager} and Eviza~\cite{setlur2016eviza} largely relied on rule-based systems, which mapped textual commands to predefined chart templates or specifications through handcrafted heuristics~\cite{de2020vismaker}. While these methods demonstrated the feasibility of automatically converting text into visualizations~\cite{moritz2018formalizing,cui2019text}, they often required extensive domain knowledge and struggled with more nuanced or ambiguous user requirements~\cite{li2021kg4vis,wang2023llm4vis}. Inspired by developments in deep learning, researchers began to incorporate neural networks to handle free-form natural language and broaden the range of supported visualization types~\cite{liu2021advisor,luo2021natural}. 

% this part should be enhanced
Building on these machine learning strategies, numerous studies have utilized LLMs to further enhance system flexibility. Recent frameworks such as Chat2VIS~\cite{maddigan2023Chat2VIS} and Prompt4Vis~\cite{liPrompt4VisPromptingLarge2024} utilize few-shot learning or query expansion to refine user queries, subsequently generating Python visualization scripts through instruction-based prompting. More recent approaches, such as MatPlotAgent~\cite{yang2024matplotagent} and PlotGen~\cite{goswamiPlotGenMultiAgentLLMbased2025}, extend these frameworks by integrating a vision-language feedback model to iteratively optimize the final code based on evaluations of the rendered visualizations. The aforementioned approaches often struggle to effectively capture user intent in complex visualization tasks. By committing to a single reasoning trajectory, they may produce code that is syntactically correct yet semantically misaligned with user expectations, requiring extensive manual adjustments. This challenge is particularly pronounced when user input is ambiguous or underspecified, leading to an iterative cycle of prompt refinement and code modification, consequently limiting the intended efficiency of automation. To address these limitations, we introduce \emph{VisPath}, a novel framework that integrates Multi-Path Reasoning with feedback from VLMs to enhance visualization code generation.

\section{Methodology}
\begin{figure*}[!htpb]
    \centering
    \includegraphics[width=1\textwidth]{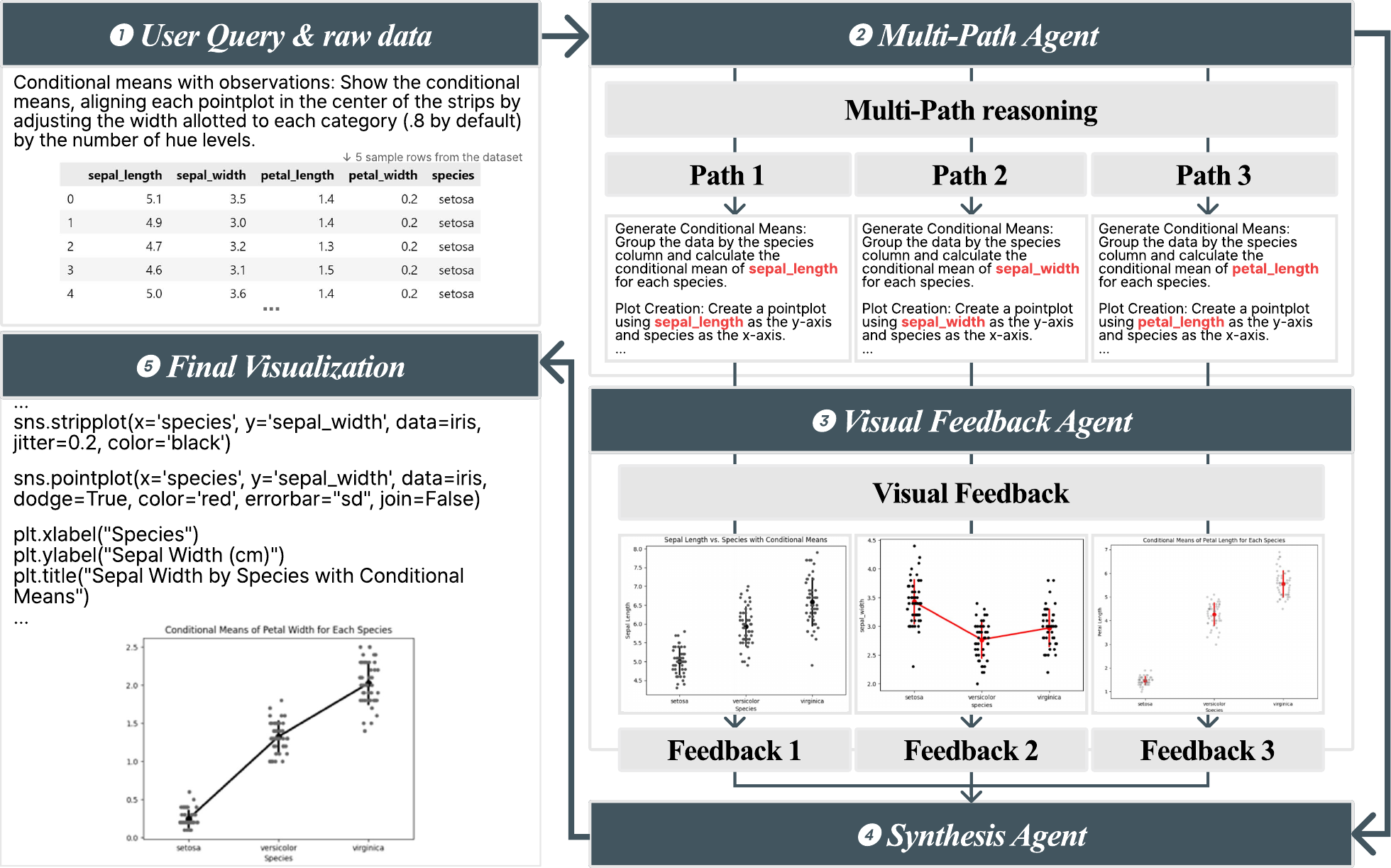}
    \caption{\textbf{Overview of the proposed \emph{VisPath} framework for creating robust visualization code generation.} The framework consists of a combination of Multi-Path Agent, Visual Feedback Agent, and Synthesis Agent.}
    \label{fig:vispath-main}
\end{figure*}

We introduce \emph{VisPath}, a framework for robust visualization code generation that combines diverse reasoning with visual feedback. 
Given a user query $Q$ and a dataset description $D$, \emph{VisPath} proceeds in three stages: (i) multi-path query expansion, (ii) visualization code synthesis and execution, and (iii) feedback-driven refinement. An overview is shown in Figure~\ref{fig:vispath-main}.

\paragraph{Notation.}
We denote any agent component by $G_{\bullet}(\,\cdot \mid S_{\bullet})$, where $S_{\bullet}$ is the \emph{system prompt} that specifies the role, constraints, and output format for that component. 
In contrast, code execution is performed by a non-Agent operator, denoted $\mathcal{E}(\cdot)$.

\subsection{Multi-Path Generation}
A single visualization request can support multiple valid interpretations depending on dataset schema, variable types, and implicit relationships. To mitigate brittle, single-assumption interpretations, \emph{VisPath} generates a diverse set of reasoning paths conditioned on the dataset context.
Specifically, a \emph{Multi-Path Agent} produces $K$ distinct reasoning pathways:
\begin{equation}\label{eq:paths-new}
\{R_1, R_2, \dots, R_K\} \sim G_{\text{mpa}}(Q, D \mid S_{\text{mpa}}).
\end{equation}
Here each $R_i$ is a structured plan (a logical blueprint) describing one plausible data-grounded interpretation of $Q$ under $D$ (e.g., chart type choice, grouping/aggregation assumptions, encoding decisions, and potential preprocessing). Conditioning on $D$ encourages paths that respect variable names, data types, and feasible visual encodings.

\subsection{Code Generation and Execution}
For each reasoning path $R_i$, a code-generation agent synthesizes an executable Python visualization script:
\begin{equation}\label{eq:code-new}
C_i \sim G_{\text{code}}(D, R_i \mid S_{\text{code}}), \quad i=1,\dots,K.
\end{equation}
The dataset description $D$ is provided explicitly to ground the script in the actual data context (e.g., column names, semantic types, and plausible transformations), reducing hallucinated variables and mismatched encodings.

Each generated script is then executed by a deterministic execution operator $\mathcal{E}$. We define the routed execution output as a tuple that jointly encodes executability and the observable outcome:
\begin{equation}\label{eq:exec-new}
Z_i := \mathcal{E}(C_i) =
\begin{cases}
\bigl(1,\, \mathrm{plot}(V_i)\bigr), & \text{if } C_i \text{ executes successfully and renders } V_i,\\
\bigl(0,\, \mathrm{err}(C_i)\bigr), & \text{otherwise},
\end{cases}
\end{equation}
for $i=1,\dots,K$, where the first element indicates executability and the second element is either the rendered plot image or the execution error message. This design avoids an explicit debugging loop while still preserving informative failure signals for downstream refinement.

\subsection{Feedback-Driven Code Optimization}
Next, a feedback agent evaluates each candidate by jointly considering the user intent and the observed execution outcome:
\begin{equation}\label{eq:feedback-new}
F_i \sim G_{\text{fb}}(Q, C_i, Z_i \mid S_{\text{fb}}), \quad i=1,\dots,K,
\end{equation}
where $F_i$ is structured feedback capturing (i) semantic alignment with $Q$, (ii) correctness with respect to the data context implied by $D$, and (iii) visual quality/readability (e.g., labeling, legends, scales, occlusion, layout).

Finally, a Synthesis Agent synthesizes a refined visualization program by aggregating the full set of candidates and their feedback:
\begin{equation}\label{eq:final-code-new}
C^{*} \sim G_{\text{syn}}\!\left(Q, D, \{(C_i, F_i)\}_{i=1}^{K} \mid S_{\text{syn}}\right).
\end{equation}
The output $C^{*}$ is optimized to be (1) executable, (2) faithful to the user intent in $Q$, and (3) visually informative given the dataset context $D$, by selectively inheriting strengths and correcting weaknesses identified across the candidate set.

\paragraph{Summary.}
\emph{VisPath} generates multiple dataset-aware reasoning paths, translates them into candidate scripts, executes each script to obtain either a plot or an error signal, and then uses structured multimodal feedback to synthesize these signals into a final robust visualization program.

\section{Experiments}

\subsection{Setup}
In this section, we detail our experimental configuration, including (1) experimental datasets, (2) model specifications, and (3) baseline methods for evaluating the performance of the proposed \emph{VisPath}.

\subsubsection{Experimental Datasets}

We evaluate our approach on two Text-to-Visualization benchmarks: \textit{MatPlotBench}~\cite{yang2024matplotagent} and \textit{Qwen-Agent Code Interpreter Benchmark}. Specifically, \textit{MatPlotBench} comprises 100 items with ground truth images; we focus on its simple instruction subset for nuanced queries. In this paper, we use \emph{underspecification} to mean that a query omits at least one decision that must be instantiated in code (e.g., plot construction primitives, data transformations, or layout conventions). Although each benchmark item provides a single reference visualization, many prompts still require such implicit choices. For example, a broken-axis plot typically entails a two-panel shared-x layout with different y-limits, and polar bar charts require a polar projection with angles in radians. VisPath is designed to surface and resolve these implicit choices via multi-path concretization. The \textit{Qwen-Agent Code Interpreter Benchmark} contains 295 records, of which 163 are visualization-related, and evaluates Python code-interpreter agents using Code Executability and Code Correctness on tasks including data visualization.
\newpage
\subsubsection{Models Used}
\paragraph{Large Language Models (LLMs)} For the code inference stage, we experiment with \textit{GPT-4o mini}~\cite{achiam2023gpt} and \textit{Gemini 2.0 Flash}~\cite{team2024gemini} to generate candidate visualization code from the reasoning paths. Both models are configured with a temperature of 0.2 to ensure precise and focused outputs, in line with previous work~\cite{yang2024matplotagent}. To evaluate the generated code quality and guide the subsequent optimization process, we utilize \textit{GPT-4o}~\cite{achiam2023gpt} and \textit{Gemini 2.0 Flash}~\cite{team2024gemini} as our visualization feedback model, which provides high-quality reference assessments.

\paragraph{Vision-Language Models (VLMs)} In order to assess the visual quality and correctness of the rendered plots, we incorporate vision evaluation models into our framework. Specifically, \textit{GPT-4o}~\cite{achiam2023gpt} is employed for detailed plot evaluation in all evaluation tasks. This setup ensures the thorough evaluation of both the syntactic correctness of the code and the aesthetic quality of the resulting visualizations.

\subsubsection{Evaluation Metrics}
In our experiments, we utilized evaluation metrics introduced by previous work to ensure consistency and comparability. \textit{MatPlotBench}~\cite{bisong2019matplotlib} assesses graph generation models using two key metrics: Plot Score, which measures similarity to the Ground Truth (0–100), and Executable Score, which represents the percentage of error-free code executions. \emph{Qwen-Agent Code Interpreter benchmark}\footnote{https://github.com/QwenLM/Qwen-Agent/blob/main/benchmark/code_interpreter/README.md} evaluates visualization models based on \textit{Visualization-Hard} and \textit{Visualization-Easy}. Compared to \textit{MatPlotBench}, \emph{Qwen-Agent Code Interpreter benchmark} assesses image alignment via a code correctness metric. Previous studies showed that GPT-based VLM evaluations align well with human assessments~\cite{yang2024matplotagent}, hence VLM was used for evaluation.

\subsubsection{Baseline Methods}
We compare \textit{VisPath} against competitive baselines: (1) \textit{Zero-Shot} directly generates visualization code without intermediate reasoning, (2) \textit{CoT Prompting} uses Chain-of-Thought (CoT) prompting to articulate its reasoning, while (3) \textit{Chat2VIS}~\cite{maddigan2023Chat2VIS} employs guiding prefixes to mitigate ambiguity, and (4) \textit{MatPlotAgent}~\cite{yang2024matplotagent} first expands the query and then refines the code via a self-debugging loop with feedback. Moreover, our proposed framework \textit{VisPath} generates three reasoning paths with corresponding visual feedback to refine the final output. For a fair comparison aligned with our experimental setting, \textit{MatPlotAgent} is limited to three iterations, and uses a critique-based debugging loop.

% ####################### %

\begin{table*}[t]
\caption{\textbf{Performance comparison of various methods across different benchmarks.} 
Visualization-Hard and Visualization-Easy refer to the Accuracy of Code Execution Results on different subsets of the \textit{Qwen-Agent Code Interpreter benchmark}. 
    \textbf{Bold text} indicates the best performance, 
    \underline{underlined text} indicates the second-best performance.
    \textsuperscript{\textdagger} denotes our proposed method.}
\renewcommand{\arraystretch}{1.2}
\centering
\resizebox{\textwidth}{!}{%
\begin{tabular}{@{}c l c c c c c@{}}
\toprule[1pt]
\multirow{2}{*}{\textbf{Model}} & \multirow{2}{*}{\textbf{Methods}} & \multicolumn{2}{c}{\textbf{MatPlotBench}} & \multicolumn{3}{c}{\textbf{Qwen-Agent Code Interpreter benchmark}} \\
\cmidrule(l){3-4} \cmidrule(l){5-7}
& & \textbf{Plot Score} & \textbf{Executable Rate} (\%) & \textbf{Visualization-Hard} & \textbf{Visualization-Easy} & \textbf{Avg.} \\
\midrule

%=========================
% gpt-4o-mini
%=========================
\multirow{5}{*}{\centering \textbf{GPT-4o mini}}
& Zero-Shot & 62.38  & 53 & 59.68 & 45.50  & 52.59 \\
& CoT Prompting & 61.95 & 50 & 57.50 & 40.00  & 48.75 \\
& Chat2VIS & 56.98 & 53 & 59.36 & 36.50 & 47.93 \\
& MatPlotAgent & \underline{63.90} & \underline{58} & \underline{67.50} & \underline{53.25} & \underline{60.38} \\
& \textbf{VisPath\textsuperscript{\textdagger} (Ours)} & \textbf{66.12} & \textbf{60} & \textbf{70.68} & \textbf{57.23} & \textbf{63.96} \\
\midrule

%=========================
% Gemini 2.0 flash
%=========================
\multirow{5}{*}{\centering \textbf{Gemini 2.0 Flash}}
& Zero-Shot & 55.00 & 54 & 68.97 & 52.18 & 60.58 \\
& CoT Prompting & 53.56  & \underline{61} & 40.00 & \textbf{63.89} & 51.95 \\
& Chat2VIS & 54.89 & 55 & 59.36 & 56.50 & 57.93 \\
& MatPlotAgent & \underline{56.31} & 58 & \underline{77.62} & 51.50 & \underline{64.56} \\
& \textbf{VisPath\textsuperscript{\textdagger} (Ours)} & \textbf{59.37} & \textbf{63} & \textbf{80.79} & \underline{57.17} & \textbf{68.98} \\
\bottomrule[1pt]
\end{tabular}%
}
    \label{tab:performance-comparison}
    \end{table*}
    
\subsection{Experimental Analysis}
\emph{VisPath} is evaluated against four baselines, as shown in Table~\ref{tab:performance-comparison}. Zero-shot prompting generates visualization code directly from natural language queries without intermediate reasoning. Despite its computational efficiency, this method frequently fails to handle ambiguity or under-specification, resulting in misaligned or incomplete outputs. On \textit{MatPlotBench (GPT-4o mini)}, it achieves a Plot Score of 62.38 and an Executable Rate of 53\%. CoT prompting further introduces a single reasoning step to expose intermediate decisions and improve interpretability. However, on \textit{MatPlotBench}, it slightly underperforms Zero-Shot in both Plot Score and Executable Rate, indicating reliance on a fixed reasoning path may reduce adaptability to diverse input structures.

\textit{Chat2VIS} extends CoT prompting by adopting prefix templates to improve coherence and reduce ambiguity in user instructions. While effective for well-structured queries, its reliance on fixed templates limits adaptability to underspecified or unconventional requests. Such limitation is evident in its performance on \textit{MatPlotBench}, where it achieves a Plot Score of 56.98 and an Executable Rate of 53\%. Furthermore, \textit{MatPlotAgent} incorporates query expansion and iterative self-debugging mechanisms to enhance robustness. While effective at correcting execution-level errors, its revisions are confined to localized adjustments and do not address higher-order semantic ambiguities. 

In contrast, our proposed framework \emph{VisPath} is specifically designed to overcome these limitations observed in prior methods by dynamically generating multiple reasoning paths and refining them through structured visual feedback. In particular, template-based approaches such as \textit{Chat2VIS} offer limited adaptability due to their reliance on predefined input formats, while methods such as \textit{MatPlotAgent} focus on localized corrections without addressing broader semantic ambiguity. Unlike prior methods, \emph{VisPath} generates diverse interpretations of user intent and evaluates them holistically using structured vision-language feedback. This enables more flexible handling of under-specified or ambiguous inputs, resulting in semantically aligned and executable visualizations. 

Evaluated across multiple benchmark settings, \emph{VisPath} notably outperforms baselines, achieving up to 9.14 point gains in Plot Score and a 10\% point increase in Executable Rate. These improvements demonstrate \emph{VisPath}’s robustness in exploring diverse reasoning paths and refining outputs through visual feedback, which reduces semantic ambiguity and enhances execution reliability.

\subsection{Ablation Study}
To further examine the robustness and design choices of \emph{VisPath}, we conduct a series of ablation experiments. Specifically, we analyze the following three aspects: (i) varying the number of generated reasoning paths, (ii) the effect of removing visual feedback during integration, and (iii) the contribution of visual feedback beyond binary executability.

\begin{figure}
    \centering
    \includegraphics[width=0.7\linewidth]{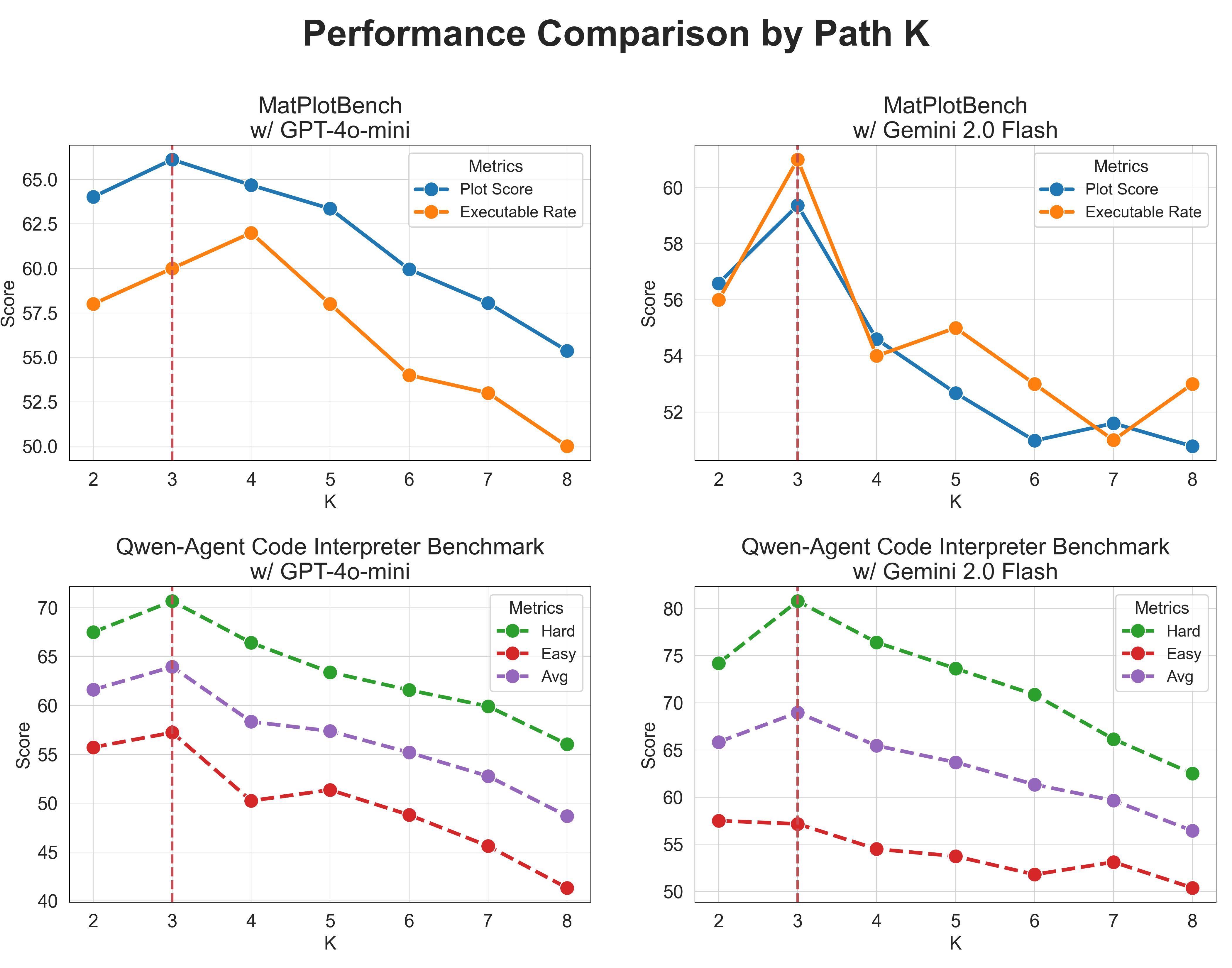}
    \caption{\textbf{Effect of varying the number of reasoning paths $K$ on performance across datasets and models.} Metrics include Plot Score, Executable Rate. The results show that $K=3$ achieves the best overall balance, with larger $K$ values reducing performance.}
    \label{fig:ablation-study-1}
\end{figure}

\subsubsection{Varying the Number of Reasoning Paths}
To investigate the contribution of reasoning path diversity, we conducted ablation experiments by varying the number of generated reasoning paths \( K \). In particular, we extended the range of \( K \) from 2 to 8 to examine the effect of increased path on the overall performance of \emph{VisPath}, as shown in Figure~\ref{fig:ablation-study-1}.

We observe a consistent pattern across all model and dataset combinations: performance improves as $K$ increases from 2 to 3, confirming that limited diversity $(K=2)$ often fails to capture nuanced interpretations of user queries. While $K=4$ achieves the highest executable rate on \textit{MatPlotBench} with GPT-4o mini (62\%), we further extend our analysis up to  $K$ = 8 to comprehensively assess the impact of reasoning path diversity. However, beyond  $K=4$, we observe diminishing returns and even performance degradation, which is likely due to noisy or redundant reasoning paths. While added diversity initially aids interpretation, excessive expansion burdens the integration process and reduces overall efficiency. 

Among all configurations tested up to $K=8$, $K=3$ emerges as the most balanced choice, offering substantial performance gains in both the Executable Rate and the Plot Score while avoiding the inefficiencies observed at higher values of $K$. Hence, we adopt $K=3$ as the default configuration throughout our experiments.

\subsubsection{Robustness with a Simple Integration}
We evaluate an alternative integration strategy that simplifies the aggregation of multiple reasoning paths to further validate the robustness of \emph{VisPath}.

\begin{table}[t]
\caption{\textbf{Performance comparison of \emph{VisPath} with and without visual feedback.}
Results are reported on \textit{MatPlotBench} (Plot Score, Executable Rate) and the average score on the \textit{Qwen-Agent Code Interpreter benchmark} for two LLMs.}
\label{tab:feedback-ablation}
\renewcommand{\arraystretch}{1.15}
\centering
\begin{tabular}{l l cc c}
\toprule
\textbf{Model} & \textbf{Setting} 
& \multicolumn{2}{c}{\textbf{MatPlotBench}} 
& \textbf{Qwen-Agent} \\
\cmidrule(lr){3-4}
& & \textbf{Plot Score} & \textbf{Exec. Rate (\%)} & \textbf{Avg.} \\
\midrule

\multirow{2}{*}{\textbf{GPT-4o mini}}
& w/o visual feedback 
& 63.76 & 56 & 58.00 \\
& \textbf{w/ visual feedback} 
& \textbf{66.12} & \textbf{60} & \textbf{63.96} \\
\midrule

\multirow{2}{*}{\textbf{Gemini 2.0 Flash}}
& w/o visual feedback 
& 55.28 & 57 & 64.03 \\
& \textbf{w/ visual feedback} 
& \textbf{59.37} & \textbf{63} & \textbf{68.98} \\
\bottomrule
\end{tabular}
\end{table}

Instead of refining each candidate visualization with VLM-based feedback, this approach aggregates three candidate codes: each derived from a distinct reasoning path, without intermediate corrections. This setup reduces computational overhead and execution time while preserving the benefits of interpretive diversity.

As shown in Table~\ref{tab:feedback-ablation}, even under this simplified configuration, \emph{VisPath} outperforms all baseline methods, confirming that Multi-Path Reasoning alone offers a strong foundation for visualization code generation. While full feedback-driven optimization leads to additional performance improvements, this result highlights that the primary strength of \emph{VisPath} lies in its capacity to explore and leverage diverse reasoning trajectories. The framework remains effective and adaptable, even with minimal refinements, further validating the importance of its core design.

\begin{table}[t]
\caption{\textbf{Ablation study isolating the impact of rendered visual feedback.}
Comparison between structured plot-based feedback and binary execution-only feedback in \emph{VisPath}.}
\label{tab:ablation-study3}
\renewcommand{\arraystretch}{1.15}
\centering
\begin{tabular}{l l cc}
\toprule
\textbf{LLM} & \textbf{Feedback Type} 
& \textbf{Plot Score} & \textbf{Exec. Rate (\%)} \\
\midrule

\multirow{2}{*}{\textbf{GPT-4o mini}}
& Binary-only feedback 
& 64.82 & 58 \\
& \textbf{Visual feedback} 
& \textbf{66.12} & \textbf{60} \\
\midrule

\multirow{2}{*}{\textbf{Gemini 2.0 Flash}}
& Binary-only feedback 
& 57.68 & 59 \\
& \textbf{Visual feedback} 
& \textbf{59.37} & \textbf{63} \\
\bottomrule
\end{tabular}
\end{table}

\subsubsection{Distinct Contribution of Visual Feedback}
To assess the role of visual feedback in improving code quality, we compare two variants of our framework. The first, \emph{VisPath (w/ feedback)}, uses a VLM to evaluate both rendered plots and error messages. The second, \emph{VisPath Execute (w/ binary feedback)}, simplifies evaluation by relying solely on the binary success or failure of code execution.

Incorporating rendered visual feedback improves the Executable Rate by $2\%-4\%$ and consistently boosts the Plot Score across both LLMs, as shown in Table~\ref{tab:ablation-study3}. 
On GPT-4o mini, Plot Score increases from 64.82 to 66.12 (+1.30) and Executable Rate from 58\% to 60\% (+2 points). On Gemini 2.0 Flash, Plot Score rises from 57.68 to 59.37 (+1.69), and Executable Rate from 59\% to 63\% (+4 points). Despite the numerical gains being modest, the results demonstrate the unique value of structured visual evaluation. Visual feedback enables more refined and user-aligned outputs by capturing subtle rendering issues that may not affect executability, demonstrating its importance in the final synthesis stage. 

\begin{figure}
    \centering
    \includegraphics[width=\linewidth]{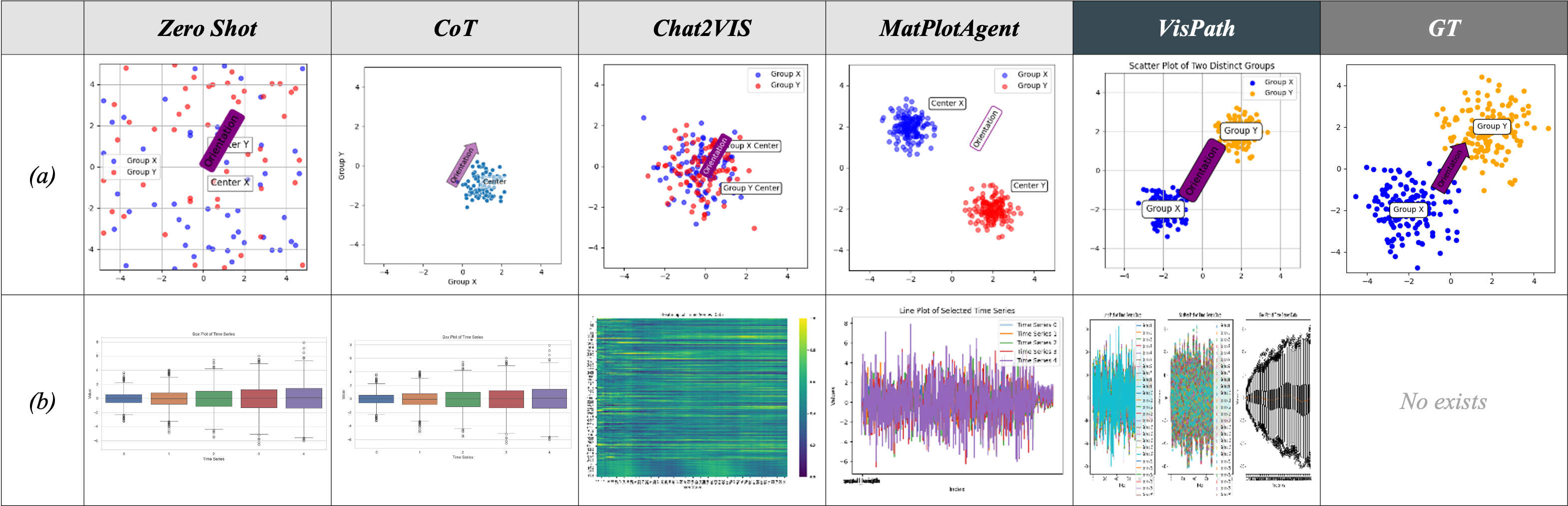}
    \caption{\textbf{(a) Scatter plot generation with explicit spatial and annotation constraints.} User query: “Create a scatter plot of two distinct sets of random data, each containing 150 points. The first set (Group X) should be centered around (-2, -2) and visualized in blue, and the second set (Group Y) should be centered around (2, 2) and visualized in orange. Label each group at their respective centers with a round white box (...) ” 
    \\ \textbf{(b) Large-scale time-series visualization task.} User query: “Visualize a large number of time series in three different ways.” This task evaluates each method’s ability to interpret an underspecified request, select appropriate plotting strategies, and compose multiple complementary visualizations for dense temporal data. \textbf{GT denotes the ground-truth visualization.}}
    \label{fig:timeseries}
\end{figure}

\section{Discussion}
\subsection{Effectiveness of Multi-Path Reasoning and Feedback-Driven Refinement}
Our proposed \emph{VisPath} framework substantially advances visualization code generation by addressing the core weaknesses of existing methods: limited interpretive flexibility and insufficient refinement. By employing Multi-Path Reasoning, \emph{VisPath} explores diverse interpretations of user intent, which leads to more accurate visualizations, especially for ambiguous queries. Experimental results confirm its superiority: \emph{VisPath} outperforms all baselines on both \textit{MatPlotBench} and the \textit{Qwen-Agent Code Interpreter benchmark}, with up to 9.14\% improvement in Plot Score and 10\% in Executable Rate. 

Also, Fig.~\ref{fig:timeseries} compares \emph{VisPath} with the baselines. VisPath matches the ground-truth structure more closely when GT is available, and still produces meaningful multi-view visualizations in Fig.~\ref{fig:timeseries}(b) where no GT exists. The results further show that \emph{VisPath} correctly follows layout constraints and handles vague time-series prompts, such as centering data groups and creating multiple views for large datasets. This suggests that the framework aligns more consistently with user intent in scenarios that remain challenging for other methods. Ablation studies further validate \emph{VisPath}'s design. First, increasing the number of reasoning paths enhances both visual quality and code executability. Second, even without visual feedback, Multi-Path Reasoning alone proves highly effective. Third, using structured plot-based feedback, rather than binary execution signals, significantly improves output alignment with user intent, confirming the value of our feedback-driven optimization module.

\subsection{Cost \& Latency Trade-offs}
To assess the latency impact of Multi-Path Reasoning and structured visual feedback, we compared the iteration counts of \emph{VisPath} against the \emph{MatPlotAgent} baseline on \emph{MatPlotBench} and the \emph{Qwen} benchmark. Iterations were measured across four categories: Query Expansion, Code Generation, Visual Feedback, and Editor. While \emph{MatPlotAgent} uses dynamic iterations based on feedback, \emph{VisPath} employs a fixed structure of $k=3$ reasoning paths for each query expansion.

\begin{table}[h]
\caption{\textbf{Iteration counts across components for \emph{MatPlotAgent} and \emph{VisPath}.} 
Benchmarks differ in size: \emph{MatPlotBench} has 100 rows, while the \emph{Qwen} benchmark has 80 rows.}
\centering
\footnotesize
\begin{adjustbox}{max width=\linewidth}
\begin{tabular}{lcccc}
\toprule
& \multicolumn{2}{c}{\emph{MatPlotBench} (100 rows)} & \multicolumn{2}{c}{\emph{Qwen} (80 rows)} \\
\cmidrule(lr){2-3} \cmidrule(lr){4-5}
& \emph{MatPlotAgent} & \emph{VisPath} & \emph{MatPlotAgent} & \emph{VisPath} \\
\midrule
Query Expansion  & 100 & 100 & 80  & 80 \\
Code Generation  & 312 & 300 & 240 & 240 \\
Visual Feedback  & 49  & 300 & 52  & 240 \\
Editor           & 285 & 100 & 226 & 80 \\
\midrule
\textbf{Total Iterations}   & 746 & 800 & 598 & 640 \\
\textbf{Avg Iterations / Row} & 7.46 & 8.00 & 7.475 & 8.00 \\
\bottomrule
\end{tabular}
\end{adjustbox}
\label{tab:cost_latency}
\end{table}

As shown in Table~\ref{tab:cost_latency}, \emph{MatPlotAgent} averaged 7.46 iterations on \emph{MatPlotBench}, while \emph{VisPath} required 8. Importantly, \emph{VisPath} incurs higher iteration counts for Visual Feedback but substantially fewer for the Editor stage (285 vs.\ 100 on \emph{MatPlotBench}; 226 vs.\ 80 on the \emph{Qwen} benchmark). Similarly, on the \emph{Qwen} benchmark, \emph{MatPlotAgent} averaged 7.475 iterations, with 8 for \emph{VisPath}. This corresponds to a marginal increase of only 0.525–0.54 iterations, despite \emph{VisPath} delivering substantial improvements in execution success and visual quality. 

The ablation study (\ref{fig:ablation-study-1}) indicate that accuracy gains do not scale proportionally with $k$, suggesting that efficacy of \emph{VisPath} is driven by its feedback-oriented design rather than brute-force iteration. By redistributing computational effort from editor retries to structured visual feedback, the system achieves meaningful performance without significant overhead. Since model calls directly impact latency and cost, $k=3$ provides the effective trade-off between robustness and efficiency for deployment beyond controlled benchmarks.

\section{Conclusion}
In this work, we present \emph{VisPath}, a framework that leverages Multi-Path Reasoning and feedback-driven optimization to enhance automated visualization code generation. Unlike prior methods, our approach seamlessly combines Multi-Path Reasoning with feedback-driven optimization. By accurately capturing diverse user intents and iteratively refining the generated code, \emph{VisPath} achieves notable improvements in both execution success and visual quality on challenging benchmarks such as \emph{MatPlotBench} and the \emph{Qwen-Agent Code Interpreter Benchmark}. By prioritizing adaptability, \emph{VisPath} is uniquely positioned to handle ambiguous user queries through a combination of diverse reasoning paths and visual feedback integration. 
Future work could explore \emph{VisPath}’s adaptability in more dynamic, real-world scenarios, further broadening its scope and practical utility in complex data analysis contexts.

\section{Limitation}
Despite its effectiveness, the current framework focuses on a feedback mechanism that assess query-code and query-plot alignment, which may overlook fine-grained elements essential to interpretability. Thus, future work could improve feedback depth by assessing individual plot components, such as readability and visual coherence, enabling more precise and refined visualization code generation. 

Moreover, while achieving strong performance, \emph{VisPath} requires several rounds of agent interaction, including multi-path reasoning, execution, and feedback integration, which may introduce inefficiencies in certain use cases. Future work could explore ways to selectively identify the most promising reasoning paths early in the process, reducing redundant computation while preserving the benefits of diverse interpretation.

Finally, our evaluation primarily focuses on quantitative metrics such as Plot Score and Executable Rate. While these metrics offer objective insights, they may not fully capture user-perceived quality, usability, or interpretability of the generated visualizations. Conducting user studies or expert reviews could provide complementary qualitative evidence and further validate the practical utility of \emph{VisPath}.

\bibliographystyle{splncs04}
\bibliography{references}

@article{maddigan2023chat2vis,
  title={Chat2vis: Fine-tuning data visualisations using multilingual natural language text and pre-trained large language models},
  author={Maddigan, Paula and Susnjak, Teo},
  journal={arXiv preprint arXiv:2303.14292},
  year={2023}
}

@article{li2024visualization,
  title={Visualization generation with large language models: An evaluation},
  author={Li, Guozheng and Wang, Xinyu and Aodeng, Gerile and Zheng, Shunyuan and Zhang, Yu and Ou, Chuangxin and Wang, Song and Liu, Chi Harold},
  journal={arXiv preprint arXiv:2401.11255},
  year={2024}
}

@article{unwin2020data,
  title={Why is data visualization important? what is important in data visualization?},
  author={Unwin, Antony},
  journal={Harvard Data Science Review},
  volume={2},
  number={1},
  pages={1},
  year={2020}
}

@article{wang2015big,
  title={Big data and visualization: methods, challenges and technology progress},
  author={Wang, Lidong and Wang, Guanghui and Alexander, Cheryl Ann},
  journal={Digital Technologies},
  volume={1},
  number={1},
  pages={33--38},
  year={2015}
}

@article{yang2024matplotagent,
  title={Matplotagent: Method and evaluation for llm-based agentic scientific data visualization},
  author={Yang, Zhiyu and Zhou, Zihan and Wang, Shuo and Cong, Xin and Han, Xu and Yan, Yukun and Liu, Zhenghao and Tan, Zhixing and Liu, Pengyuan and Yu, Dong and others},
  journal={arXiv preprint arXiv:2402.11453},
  year={2024}
}

@article{ge2023automatic,
  title={Automatic Data Visualization Generation from Chinese Natural Language Questions},
  author={Ge, Yan and Wei, Victor Junqiu and Song, Yuanfeng and Zhang, Jason Chen and Wong, Raymond Chi-Wing},
  journal={arXiv preprint arXiv:2309.07650},
  year={2023}
}

@inproceedings{zhang2024gpt,
  title={Is gpt-4v (ision) all you need for automating academic data visualization? exploring vision-language models’ capability in reproducing academic charts},
  author={Zhang, Zhehao and Ma, Weicheng and Vosoughi, Soroush},
  booktitle={Findings of the Association for Computational Linguistics: EMNLP 2024},
  pages={8271--8288},
  year={2024}
}

@inproceedings{sharif2024understanding,
  title={Understanding and reducing the challenges faced by creators of accessible online data visualizations},
  author={Sharif, Ather and Kim, Joo Gyeong and Xu, Jessie Zijia and Wobbrock, Jacob O},
  booktitle={Proceedings of the 26th International ACM SIGACCESS Conference on Computers and Accessibility},
  pages={1--20},
  year={2024}
}

@article{bresciani2015pitfalls,
  title={The pitfalls of visual representations: A review and classification of common errors made while designing and interpreting visualizations},
  author={Bresciani, Sabrina and Eppler, Martin J},
  journal={Sage Open},
  volume={5},
  number={4},
  pages={2158244015611451},
  year={2015},
  publisher={SAGE Publications Sage CA: Los Angeles, CA}
}

@article{xie2024haichart,
  title={HAIChart: Human and AI Paired Visualization System},
  author={Xie, Yupeng and Luo, Yuyu and Li, Guoliang and Tang, Nan},
  journal={arXiv preprint arXiv:2406.11033},
  year={2024}
}

@article{dibia2019data2vis,
  title={Data2vis: Automatic generation of data visualizations using sequence-to-sequence recurrent neural networks},
  author={Dibia, Victor and Demiralp, {\c{C}}a{\u{g}}atay},
  journal={IEEE computer graphics and applications},
  volume={39},
  number={5},
  pages={33--46},
  year={2019},
  publisher={IEEE}
}

@inproceedings{qian2021learning,
  title={Learning to recommend visualizations from data},
  author={Qian, Xin and Rossi, Ryan A and Du, Fan and Kim, Sungchul and Koh, Eunyee and Malik, Sana and Lee, Tak Yeon and Chan, Joel},
  booktitle={Proceedings of the 27th ACM SIGKDD conference on knowledge discovery \& data mining},
  pages={1359--1369},
  year={2021}
}

@article{wang2023data,
  title={Data Formulator: AI-powered Concept-driven Visualization Authoring},
  author={Wang, Chenglong and Thompson, John and Lee, Bongshin},
  journal={IEEE Transactions on Visualization and Computer Graphics},
  year={2023},
  publisher={IEEE}
}

@inproceedings{vondrick2013hoggles,
  title={Hoggles: Visualizing object detection features},
  author={Vondrick, Carl and Khosla, Aditya and Malisiewicz, Tomasz and Torralba, Antonio},
  booktitle={Proceedings of the IEEE International Conference on Computer Vision},
  pages={1--8},
  year={2013}
}

@article{demiralp2017foresight,
  title={Foresight: Recommending visual insights},
  author={Demiralp, {\c{C}}a{\u{g}}atay and Haas, Peter J and Parthasarathy, Srinivasan and Pedapati, Tejaswini},
  journal={arXiv preprint arXiv:1707.03877},
  year={2017}
}

@article{saket2018task,
  title={Task-based effectiveness of basic visualizations},
  author={Saket, Bahador and Endert, Alex and Demiralp, {\c{C}}a{\u{g}}atay},
  journal={IEEE transactions on visualization and computer graphics},
  volume={25},
  number={7},
  pages={2505--2512},
  year={2018},
  publisher={IEEE}
}

@article{xiao2023let,
  title={Let the chart spark: Embedding semantic context into chart with text-to-image generative model},
  author={Xiao, Shishi and Huang, Suizi and Lin, Yue and Ye, Yilin and Zeng, Wei},
  journal={IEEE Transactions on Visualization and Computer Graphics},
  year={2023},
  publisher={IEEE}
}

@article{han2023chartllama,
  title={Chartllama: A multimodal llm for chart understanding and generation},
  author={Han, Yucheng and Zhang, Chi and Chen, Xin and Yang, Xu and Wang, Zhibin and Yu, Gang and Fu, Bin and Zhang, Hanwang},
  journal={arXiv preprint arXiv:2311.16483},
  year={2023}
}

@inproceedings{barrett2005matplotlib,
  title={matplotlib--A Portable Python Plotting Package},
  author={Barrett, Paul and Hunter, John and Miller, J Todd and Hsu, J-C and Greenfield, Perry},
  booktitle={Astronomical data analysis software and systems XIV},
  volume={347},
  pages={91},
  year={2005}
}

@article{bisong2019matplotlib,
  title={Matplotlib and seaborn},
  author={Bisong, Ekaba and Bisong, Ekaba},
  journal={Building machine learning and deep learning models on google cloud platform: A comprehensive guide for beginners},
  pages={151--165},
  year={2019},
  publisher={Springer}
}

@book{zhu2013data,
  title={Data visualization with D3. js cookbook},
  author={Zhu, Nick Qi},
  year={2013},
  publisher={Packt Publishing Ltd}
}

@article{chen2022type,
  title={Type-directed synthesis of visualizations from natural language queries},
  author={Chen, Qiaochu and Pailoor, Shankara and Barnaby, Celeste and Criswell, Abby and Wang, Chenglong and Durrett, Greg and Dillig, I{\c{s}}il},
  journal={Proceedings of the ACM on Programming Languages},
  volume={6},
  number={OOPSLA2},
  pages={532--559},
  year={2022},
  publisher={ACM New York, NY, USA}
}

@article{zhang2024chartifytext,
  title={ChartifyText: Automated Chart Generation from Data-Involved Texts via LLM},
  author={Zhang, Songheng and Wang, Lei and Li, Toby Jia-Jun and Shen, Qiaomu and Cao, Yixin and Wang, Yong},
  journal={arXiv preprint arXiv:2410.14331},
  year={2024}
}

@article{chen2022nl2interface,
  title={Nl2interface: Interactive visualization interface generation from natural language queries},
  author={Chen, Yiru and Li, Ryan and Mac, Austin and Xie, Tianbao and Yu, Tao and Wu, Eugene},
  journal={arXiv preprint arXiv:2209.08834},
  year={2022}
}

@inproceedings{rashid2022text2chart,
  title={Text2chart: A multi-staged chart generator from natural language text},
  author={Rashid, Md Mahinur and Jahan, Hasin Kawsar and Huzzat, Annysha and Rahul, Riyasaat Ahmed and Zakir, Tamim Bin and Meem, Farhana and Mukta, Md Saddam Hossain and Shatabda, Swakkhar},
  booktitle={Pacific-Asia Conference on Knowledge Discovery and Data Mining},
  pages={3--16},
  year={2022},
  organization={Springer}
}

@inproceedings{wu2022nuwa,
  title={N{\"u}wa: Visual synthesis pre-training for neural visual world creation},
  author={Wu, Chenfei and Liang, Jian and Ji, Lei and Yang, Fan and Fang, Yuejian and Jiang, Daxin and Duan, Nan},
  booktitle={European conference on computer vision},
  pages={720--736},
  year={2022},
  organization={Springer}
}

@article{wongsuphasawat2015voyager,
  title={Voyager: Exploratory analysis via faceted browsing of visualization recommendations},
  author={Wongsuphasawat, Kanit and Moritz, Dominik and Anand, Anushka and Mackinlay, Jock and Howe, Bill and Heer, Jeffrey},
  journal={IEEE transactions on visualization and computer graphics},
  volume={22},
  number={1},
  pages={649--658},
  year={2015},
  publisher={IEEE}
}

@article{li2021kg4vis,
  title={KG4Vis: A knowledge graph-based approach for visualization recommendation},
  author={Li, Haotian and Wang, Yong and Zhang, Songheng and Song, Yangqiu and Qu, Huamin},
  journal={IEEE Transactions on Visualization and Computer Graphics},
  volume={28},
  number={1},
  pages={195--205},
  year={2021},
  publisher={IEEE}
}

@article{wang2023llm4vis,
  title={LLM4Vis: Explainable visualization recommendation using ChatGPT},
  author={Wang, Lei and Zhang, Songheng and Wang, Yun and Lim, Ee-Peng and Wang, Yong},
  journal={arXiv preprint arXiv:2310.07652},
  year={2023}
}

@inproceedings{setlur2016eviza,
  title={Eviza: A natural language interface for visual analysis},
  author={Setlur, Vidya and Battersby, Sarah E and Tory, Melanie and Gossweiler, Rich and Chang, Angel X},
  booktitle={Proceedings of the 29th annual symposium on user interface software and technology},
  pages={365--377},
  year={2016}
}

@article{cui2019text,
  title={Text-to-viz: Automatic generation of infographics from proportion-related natural language statements},
  author={Cui, Weiwei and Zhang, Xiaoyu and Wang, Yun and Huang, He and Chen, Bei and Fang, Lei and Zhang, Haidong and Lou, Jian-Guan and Zhang, Dongmei},
  journal={IEEE transactions on visualization and computer graphics},
  volume={26},
  number={1},
  pages={906--916},
  year={2019},
  publisher={IEEE}
}

@article{moritz2018formalizing,
  title={Formalizing visualization design knowledge as constraints: Actionable and extensible models in draco},
  author={Moritz, Dominik and Wang, Chenglong and Nelson, Greg L and Lin, Halden and Smith, Adam M and Howe, Bill and Heer, Jeffrey},
  journal={IEEE transactions on visualization and computer graphics},
  volume={25},
  number={1},
  pages={438--448},
  year={2018},
  publisher={IEEE}
}

@article{de2020vismaker,
  title={Vismaker: a question-oriented visualization recommender system for data exploration},
  author={de Ara{\'u}jo Lima, Raul and Diniz Junqueira Barbosa, Simone},
  journal={arXiv e-prints},
  pages={arXiv--2002},
  year={2020}
}

@inproceedings{liu2021advisor,
  title={Advisor: Automatic visualization answer for natural-language question on tabular data},
  author={Liu, Can and Han, Yun and Jiang, Ruike and Yuan, Xiaoru},
  booktitle={2021 IEEE 14th Pacific Visualization Symposium (PacificVis)},
  pages={11--20},
  year={2021},
  organization={IEEE}
}

@article{luo2021natural,
  title={Natural language to visualization by neural machine translation},
  author={Luo, Yuyu and Tang, Nan and Li, Guoliang and Tang, Jiawei and Chai, Chengliang and Qin, Xuedi},
  journal={IEEE Transactions on Visualization and Computer Graphics},
  volume={28},
  number={1},
  pages={217--226},
  year={2021},
  publisher={IEEE}
}

@article{achiam2023gpt,
  title={Gpt-4 technical report},
  author={Achiam, Josh and Adler, Steven and Agarwal, Sandhini and Ahmad, Lama and Akkaya, Ilge and Aleman, Florencia Leoni and Almeida, Diogo and Altenschmidt, Janko and Altman, Sam and Anadkat, Shyamal and others},
  journal={arXiv preprint arXiv:2303.08774},
  year={2023}
}

@misc{liPrompt4VisPromptingLarge2024,
  title = {{{Prompt4Vis}}: {{Prompting Large Language Models}} with {{Example Mining}} and {{Schema Filtering}} for {{Tabular Data Visualization}}},
  author = {Li, Shuaimin and Chen, Xuanang and Song, Yuanfeng and Song, Yunze and Zhang, Chen},
  year = {2024},
  month = jan,
  number = {arXiv:2402.07909},
  eprint = {2402.07909},
  primaryclass = {cs},
  publisher = {arXiv},
  doi = {10.48550/arXiv.2402.07909},
  archiveprefix = {arXiv}
}

@misc{goswamiPlotGenMultiAgentLLMbased2025,
  title = {{{PlotGen}}: {{Multi-Agent LLM-based Scientific Data Visualization}} via {{Multimodal Feedback}}},
  author = {Goswami, Kanika and Mathur, Puneet and Rossi, Ryan and Dernoncourt, Franck},
  year = {2025},
  month = feb,
  number = {arXiv:2502.00988},
  eprint = {2502.00988},
  primaryclass = {cs},
  publisher = {arXiv},
  doi = {10.48550/arXiv.2502.00988},
  archiveprefix = {arXiv}
}

@article{team2024gemini,
  title={Gemini 1.5: Unlocking multimodal understanding across millions of tokens of context},
  author={Team, Gemini and Georgiev, Petko and Lei, Ving Ian and Burnell, Ryan and Bai, Libin and Gulati, Anmol and Tanzer, Garrett and Vincent, Damien and Pan, Zhufeng and Wang, Shibo and others},
  journal={arXiv preprint arXiv:2403.05530},
  year={2024}
}

@article{wen2025exploring,
  title={Exploring multimodal prompt for visualization authoring with large language models},
  author={Wen, Zhen and Weng, Luoxuan and Tang, Yinghao and Zhang, Runjin and Liu, Yuxin and Pan, Bo and Zhu, Minfeng and Chen, Wei},
  journal={arXiv preprint arXiv:2504.13700},
  year={2025}
}

\end{document}